\documentclass{article}
\usepackage{microtype}
\usepackage{graphicx}
\usepackage{subcaption}
\usepackage{booktabs}
\usepackage{xspace}
\usepackage{hyperref}
\usepackage{svg}
\usepackage{amsmath}

\usepackage[accepted]{mlsys2025}
\usepackage{url}

\usepackage{enumitem}
\setlist[itemize]{noitemsep, topsep=0pt, leftmargin=*}
\setlist[enumerate]{noitemsep, topsep=0pt, leftmargin=*}

\usepackage{color, xcolor}
\usepackage[colorinlistoftodos,prependcaption,textsize=tiny]{todonotes}

\newcommand{\ie}{i.e.\xspace}
\newcommand{\eg}{e.g.\xspace}

\newcommand{\para}{\noindent\textbf}

\newcommand{\presec}{\vspace{-0.08in}}
\newcommand{\postsec}{\vspace{-0.05in}}
\newcommand{\presub}{\vspace{-0.02in}}
\newcommand{\postsub}{\vspace{-0.00in}}
\newcommand{\postfig}{\vspace{-0.10in}}
\newcommand{\postfigcaption}{\vspace{-0.10in}}

\newcommand{\sysname}{\textsc{Neo}\xspace}

\mlsystitlerunning{\sysname: Saving GPU Memory Crisis with CPU Offloading for Online LLM Inference}

\begin{document}

\twocolumn[
\mlsystitle{\sysname: Saving GPU Memory Crisis with CPU Offloading\\ for Online LLM Inference}

\mlsyssetsymbol{equal}{*}

\begin{mlsysauthorlist}
\mlsysauthor{Xuanlin Jiang*}{PKU}
\mlsysauthor{Yang Zhou}{UCB,UCD}
\mlsysauthor{Shiyi Cao}{UCB}
\mlsysauthor{Ion Stoica}{UCB}
\mlsysauthor{Minlan Yu}{Harvard}
\end{mlsysauthorlist}

\mlsysaffiliation{PKU}{Peking University}
\mlsysaffiliation{UCB}{UC Berkeley}
\mlsysaffiliation{UCD}{UC Davis}
\mlsysaffiliation{Harvard}{Harvard University}

\mlsyscorrespondingauthor{Xuanlin Jiang}{xljiang@stu.pku.edu.cn}
\mlsyscorrespondingauthor{Yang Zhou}{yangzhou.rpc@gmail.com}
\mlsyskeywords{Machine Learning, MLSys}

\vskip 0.3in

\begin{abstract}
Online LLM inference powers many exciting applications such as intelligent chatbots and autonomous agents. 
Modern LLM inference engines widely rely on request batching to improve inference throughput, aiming to make it cost-efficient when running on expensive GPU accelerators. 
However, the limited GPU memory has largely limited the batch size achieved in practice, leaving significant GPU compute resources wasted. 

We present \sysname, an online LLM inference system that offloads part of attention compute and KV cache states from the GPU to the local host CPU, effectively increasing the GPU batch size and thus inference throughput. 
To this end, \sysname proposes asymmetric GPU-CPU pipelining and load-aware scheduling to balance GPU and CPU loads and fully utilize their compute and memory resources. 
We evaluate \sysname on a wide range of workloads (\ie, code generation, text summarization), GPUs (\ie, T4, A10G, H100), and LLM models (\ie, 7B, 8B, 70B). 
\sysname achieves up to 7.5$\times$, 26\%, and 14\% higher throughput compared to GPU-only approach on T4, A10G, and H100 GPUs, respectively, while maintaining the same latency; 
with more powerful CPUs, \sysname achieves up to 79.3\% throughput gain on A10G GPU. 
\end{abstract}
]

\renewcommand{\mlsysEqualContribution}{\textsuperscript{*}Work done during an internship at Harvard University.}

\printAffiliationsAndNotice{\mlsysEqualContribution}

\presec
\section{Introduction}
\postsec
\label{introduction}

The advent of auto-regressive transformer-based large language models (LLMs) has significantly reshaped existing technologies such as search engines and chatbots and empowered various new ones, such as autonomous agents and programming assistants. 
In these online scenarios, LLM inference is directly user-facing and thus requires low latency for immersive interaction; it also desires high throughput, typically via request batching, to efficiently leverage expensive hardware accelerators like GPUs~\cite{vllm}. 

However, to achieve large batch sizes for high throughput, online LLM inference requires huge GPU memory, but GPUs have limited memory resources. 
Nowadays LLM models have billions of parameters (\eg, 70B LLaMa-3.1 model~\cite{llama3}) that occupy dozens to hundreds of GB GPU memory; modern LLM inference engines like vLLM~\cite{vllm} additionally store KV cache in the GPU memory to reuse previous computations, whose size increases linearly with prompt and output length. 
As a result, the memory-bounded LLM inference workloads have created the GPU memory crisis where people demand expensive high-end GPUs with large memory sizes. 

Prior work has recognized this problem and proposed various solutions. 
One line of work is on model quantization~\cite{atom, attention_sink} to reduce model memory consumption. However, they come at the cost of lower accuracy. 
Another line of work is offloading model weights, KV cache, and compute to the CPU, such as FlexGen~\cite{flexgen}, PowerInfer~\cite{powerinfer}, TwinPilots~\cite{twinpilots}, HeteGen~\cite{hetegen}, and FastDecode~\cite{fastdecode}. 
Memory offloading could increase request batch sizes on the GPU, potentially increasing overall inference throughput; 
compute offloading avoids repetitively swapping the KV cache between the GPU and CPU, thus preventing PCIe bandwidth from becoming the bottleneck. 
Unfortunately, most of these work trades inference latency for throughput by using huge GPU batch sizes and layer-by-layer swapping~\cite{flexgen}, thus not suitable for online inference; 
the exception is FastDecode, but it uses 8 32-core AMD Epyc CPUs in remote servers to handle the offloaded compute of only one A10G GPU, where the CPUs cost much more than the GPU. 

This paper aims to achieve higher throughput for online LLM inference \textit{without compromising accuracy or latency} in a cost-efficient way---only using \textit{local host CPUs} as the offloading target that comes with the GPU ``as free''. 
Achieving this goal faces two main challenges. 
First, within each inference iteration, how to balance the compute happening on GPU and CPU to fully utilize their compute and memory resources, while not overloading them? 
This is challenging because of the vastly different characteristics between GPUs and CPUs. For example, a low-end GPU could easily have nearly TB/s memory bandwidth and hundreds of TFLOPS compute, while a high-end CPU server has only a few hundreds of GB/s memory bandwidth and few TFLOPS compute~\cite{fastdecode}. 
Prior work like FastDecode fully offloads the KV cache and attention computation to the CPU, making the CPU severely bottleneck the system. 
Second, across inference iterations, how to adaptively decide the offloading policy for real-world workloads with dynamic input/output lengths? 
Prior work like FlexGen and FastDecode assumes fixed input/output lengths across requests, and leverages static optimal offloading policies that were determined by one-time offline profiling. 
As inference iterations proceed, dynamic input/output lengths of requests in real-world workloads would easily break the optimality of such policies. 

To address these two challenges, we present \sysname\footnote{Neo is the protagonist in \textit{The Matrix} who saves humankind.} with two key designs: \textit{asymmetric GPU-CPU pipelining} within each inference iteration, and \textit{load-aware scheduling} across iterations. 
Asymmetric pipelining runs two asymmetric sub-batches concurrently: one offloads the decoding attention computation and KV cache of a \textit{subset} of requests into the CPU, and another one runs the rest in the GPU; these two sub-batches overlap with each other to balance GPU and CPU loads. 
This partial offloading will help keep the offloaded compute and memory bandwidth consumption within CPU capacity. 
Load-aware scheduling online monitors various request waiting and running queues (for the CPU and GPU), then dynamically decides which requests should be offloaded to the CPU and how to form request batches, based on principled heuristics to optimize for the highest throughput. 

Perhaps the closest work to \sysname is FastDecode, but \sysname differs in several important designs: 1) it features partial offloading to limit the offloaded computation and memory bandwidth consumption without overloading the CPU, 2) it adaptively finds the optimal point to balance GPU and CPU loads for dynamically-changing real-world workloads. 

\sysname is implemented based on SwiftLLM~\cite{swiftllm}, a simplified copy of vLLM with similar performance, but could be adapted to other frameworks like vLLM~\cite{vllm} and SGLang~\cite{sglang}. It leverages the Intel ISPC compiler~\cite{ispc} to generate efficient CPU kernels for attention computations. 
We thoroughly evaluate \sysname on AWS GPU instances (\texttt{g4} with a T4 GPU and \texttt{g5} with an A10G GPU) and our local 8$\times$H100 testbed, both with only the host CPU and memory for offloading. 
Our evaluation covers two public online inference datasets and three popular LLM models ranging from 7B, 8B, to 70B. 
\sysname achieves up to 14\%-7.5$\times$ (depending on GPUs) higher throughput over the non-offloaded version while keeping the same inference latency; it further achieves up to 79.3\% performance gains with more power CPUs. 

To the best of our knowledge, \textit{\sysname is the first CPU offloading system for online LLM inference that achieves performance gains over GPU-only systems with the same hardware cost and inference accuracy. }
We hope this opens a new door for cost-efficient LLM inference research.

\presec
\section{Background and Motivation}
\postsec
\label{sec:background}

\subsection{LLM Inference and Performance Bottleneck}
\begin{figure*}[ht]
    \centering
    \includegraphics[width=0.8\linewidth]{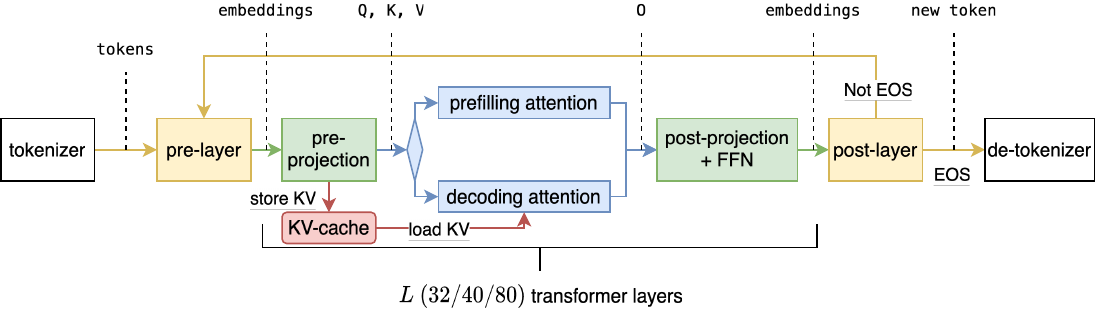}
    \vspace{-0.15in}
    \caption{Workflow of transformer-based LLM Inference.}
    \vspace{-0.15in}
    \label{fig:general_inference}
\end{figure*}

\postsub
\label{ssec:geninfer}
Auto-regressive transformer-based LLMs~\cite{attention} perceive tokens as the basic elements of languages, with the main task of predicting the next token of the given sequence. Formally, given $[t_1, t_2, \dots, t_n]$ as input, for each token $x$ in the vocabulary, an LLM needs to return the probability that $x$ is the next token of the sequence, i.e. $\Pr[t_{n+1}=x\mid t_1,\dots, t_n]$. 
To do this prediction, as illustrated in Figure~\ref{fig:general_inference}, the transformer-based LLM first converts tokens to embedding vectors. These vectors are then passed through a series of primary blocks called transformer layers, leveraging the attention mechanism~\cite{attention}. The embedding vectors remain the same shape but become more precise and context-aware after passing through each transformer layer, before finally getting through a final fully connected layer that converts embedding vectors to corresponding probabilities for each possible token. 

One inference request typically consists of prefilling and decoding stages and heavily involves the KV cache on GPU memory to reuse previous computations. 
The prefilling stage generates the initial KV cache after consuming all input tokens, while the decoding stage repetitively read-and-appends the KV cache and auto-regressively generates output tokens until an EOS (End Of Sequence). 
To optimize this process, modern LLM inference engines like vLLM~\cite{vllm} leverage iteration-level scheduling~\cite{orca} to accommodate various input/output lengths of requests, selective batching~\cite{orca} to increase performance by batching matrix multiplications, and paged attention to efficiently manage GPU memory. 

The throughput of LLM inference highly depends on the batch size the engine can achieve, which is essentially bounded by the GPU memory size due to the large KV cache~\cite{vllm}. 
Prior work~\cite{fastdecode, nanoflow} has demonstrated that the throughput would increase almost linearly as the batch size increases up to hundreds on modern A10, V100, and H100 GPUs; meanwhile, unfortunately, these GPUs fail to accommodate hundreds or even dozens of batch size, leaving significant GPU compute resources unutilized. 

\presub
\subsection{CPU Memory as a Possible Rescue}
\postsub
\label{ssec:cpu_as_rescue}

One approach to increase batch size is storing the overflowing KV cache in the main memory of the CPU, transferring it to the GPU when needed, and transferring it back when not needed. However, such repetitive KV cache swap-ins and -outs make the system severely bounded by GPU-CPU PCIe bandwidth, as shown by prior work~\cite{flexgen}. 
Fortunately, only the decoding-phase attention operation relies on the KV cache, therefore offloading this part of computation to the CPU would help avoid repetitively transferring the KV cache between the GPU and CPU. 
Moreover, this operation only takes a tiny proportion of computation (compared to other parts in the transformer architecture) and doesn't require loading model weights. 

The decoding attention operation is memory-bandwidth-bounded on both GPU and CPU due to low arithmetic intensity (\ie, FLOP per memory load)~\cite{fastdecode, nanoflow}. 
The memory bandwidth gap between GPUs and CPUs is much smaller than their compute gap. 
For example, an A10G GPU features 600 GB/s memory bandwidth and 125 TFLOPS, while a modern x86 server has around 200 GB/s memory bandwidth with 1.2 TFLOPS~\cite{fastdecode, nanoflow}; 
modern ARM processors like AWS Graviton4 offer 537.6 GB/s memory bandwidth per socket ~\cite{graviton4}. 
Therefore, although there may seem to be a huge compute gap between GPUs and CPUs (\ie, 125 vs. 1.2 TFLOPS), the actual performance gap for the decoding attention operation is relatively small because their memory bandwidths are closer (\ie, 600 vs. 200 GB/s). 

\presub
\subsection{Challenges}
\postsub
\label{ssec:challenges}

Despite being promising, there are several challenges when building an efficient LLM inference system that offloads decoding attention compute and memory to CPUs. 
These challenges stem from the fundamentally different capabilities between GPUs and CPUs (in terms of memory and compute power), and get amplified in real-world dynamically-changing inference workloads (\eg, various input/output lengths). 

\para{Challenge \#1:} How to efficiently overlap the GPU and CPU \textit{within each inference iteration}? 
GPUs have more compute and memory bandwidth but are limited in memory size, while CPUs have more memory and a decent amount of memory bandwidth, but lack strong compute power. 
Therefore, we must carefully restructure the pipeline of LLM inference to fit different pipeline modules into the right hardware, while not overloading any hardware. 
Such restructuring also needs to take care of the complex inter- and intra-transformer-layer data dependencies, without breaking transformer semantics. 

\para{Challenge \#2:} How to schedule inference requests to the GPU and CPU \textit{across inference iterations} to maintain high performance in dynamically-changing workloads? 
Prior work like FastDecode~\cite{fastdecode}, FlexGen~\cite{flexgen}, and more~\cite{attention_offloading} only consider idealistic settings where request input and output lengths are fixed, and adopt a static scheduling policy (\eg, obtained from offline profiling) to assign requests across the GPU and CPU. 
However, in real-world dynamic settings, vastly different input/output lengths would make static scheduling no longer work efficiently; instead, it would require an adaptive scheduling policy to determine the best request assignments at the per-iteration level. 

\presec
\section{\sysname Design}
\postsec
\label{sec:design}

\begin{figure}[ht]
    \centering
    \includegraphics[width=\linewidth]{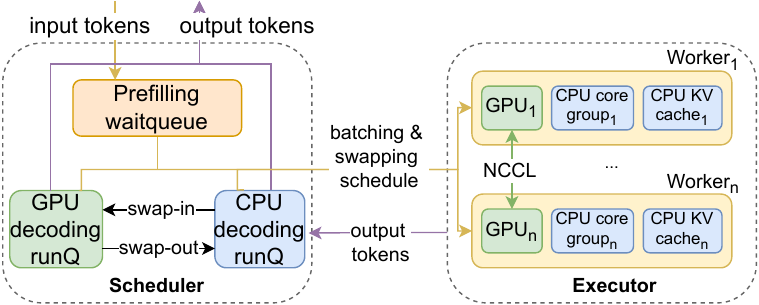}
    \vspace{-0.22in}
    \caption{Overall architecture of \sysname. ``runQ'' means ``runqueue''. }
    \vspace{-0.08in}
    \label{fig:arch}
\end{figure}

\begin{figure*}[!t]
    \centering
    \includegraphics[width=0.8\linewidth]{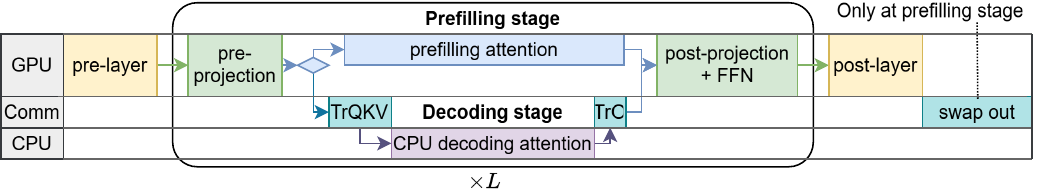}
    \vspace{-0.10in}
    \caption{Simple offloading strawman offloads all requests' KV cache and decoding attention computation to the CPU. ``Comm'' stands for GPU-CPU communication; ``TrQKV'' means transferring Q,K,V tensors to CPU; ``TrO'' means transferring attention output to GPU.}
    \vspace{-0.05in}
    \label{fig:naive-offl}
\end{figure*}

\begin{figure*}[!t]
    \centering
    \includegraphics[width=0.95\linewidth]{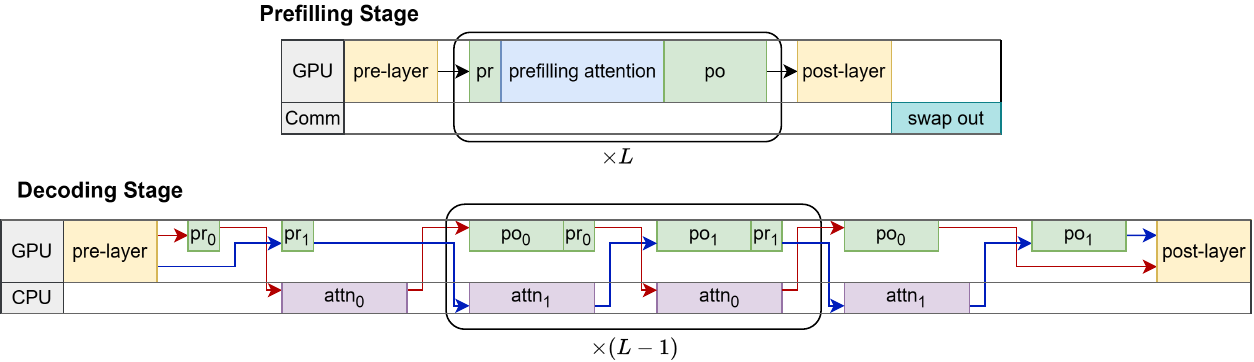}
    \vspace{-0.10in}
    \caption{Symmetric pipelining strawman forms two identical sub-batches and overlaps linear and attention operations for the decoding stage. 
    The red and blue arrows depict the data flows of the two sub-batches. 
    ``pr'' means pre-projection and ``po'' means post-projection + FFN operations; together they form the linear stage. 
    ``attn'' means attention operations. ``TrQKV''s and ``TrO''s are omitted for simplicity.
    }
    \vspace{-0.05in}
    \label{fig:naive-itlv}
\end{figure*} 

Figure~\ref{fig:arch} shows the high-level architecture of \sysname. 
\sysname consists of a request scheduler running on the CPU that maintains a prefilling waitqueue, a GPU decoding runqueue, and a CPU decoding runqueue; this scheduler makes iteration-level adaptive scheduling decisions on whether to run the incoming requests on GPU or CPU. 
\sysname features two key techniques: 1) asymmetric pipelining to fully leverage the compute resource of both GPU and CPU without overloading them (\S\ref{ssec:asymm}), and 2) load-aware scheduling to handle dynamically-changing workloads, \eg, different input/output lengths across requests (\S\ref{ssec:scheduler}). 

\presub
\subsection{Asymmetric Pipelining}
\postsub
\label{ssec:asymm}

\sysname proposes \textit{asymmetric pipelining} to address the GPU-CPU overlapping challenge in \S\ref{ssec:challenges}. 
This design offloads decoding attention (both its compute and KV cache) of a selected portion of inference requests to the CPU. 
It forms two sub-batches of requests---one mostly runs on the GPU while another runs across the GPU and CPU, and overlaps these two sub-batches to balance the load of GPU and CPU.

To motivate this design, we first explore a simple strawman called \textit{simple offloading} that offloads compute and KV cache to the CPU but leaves the GPU idle. 
Next, we examine a more intricate strawman called \textit{symmetric pipelining} used by prior work~\cite{fastdecode, attention_offloading} that tends to leave the GPU idle. 
Finally, we will arrive at our design of asymmetric pipelining, and show how it effectively addresses the issues in the simple offloading and symmetric pipelining designs. 

\para{Strawman \#1: simple offloading. } As shown in Figure~\ref{fig:naive-offl}, this design extracts the decoding attention and offloads its compute and KV cache to the CPU, while leaving the rest to the GPU. 
The rest includes prefilling attention and token-wise independent operations---referred to as the linear operations that mainly involve matrix multiplications. 
However, during these linear operations, the CPU always remains idle. 
As a result, this design fails to leverage the compute and memory resources of CPUs. 

\para{Strawman \#2: symmetric pipelining. } A straightforward approach to reducing the CPU idle time is to evenly split a single decoding batch into two sub-batches, and overlap their linear operations (on the GPU) and attention operations (on the CPU), as shown in Figure~\ref{fig:naive-itlv}. For the prefilling batch, symmetric pipelining just runs it on the GPU without offloading, as the prefilling stage requires high computation for matrix multiplication while not consuming much memory~\cite{nanoflow}. 
The output of the symmetric pipelining, \ie, the KV cache, will be swapped out to the CPU for offloading. 
This technique has been used by prior work like FastDecode~\cite{fastdecode} and more~\cite{attention_offloading}. 
Nevertheless, seemingly efficient, this design suffers from three major issues. 

\begin{figure*}[!t]
    \centering
    \includegraphics[width=0.95\linewidth]{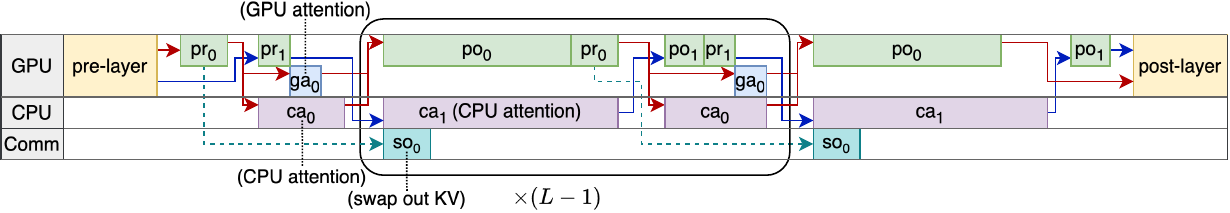}
    \vspace{-0.10in}
    \caption{Asymmetric pipelining integrates the prefilling stage into one sub-batch (red arrows) and most of the decoding attention operations into another (blue arrows). 
    ``pr'' means pre-projection, while ``po'' means post-projection + FFN operations; ``attn'' means attention operations; ``Comm'' stands for GPU-CPU communication. 
    }
    \vspace{-0.05in}
    \label{fig:asymm-itlv}
\end{figure*}

\begin{itemize}
\item Firstly, this design results in significant underutilization of GPU memory. In this scheme, the GPU solely retains the model weights and runtime activations, while the rest of the memory---which stores the KV cache in non-offloading settings---remains unused. 

\item Secondly, this design fails to achieve balanced GPU-CPU overlapping.
This is because 1) it entirely overlooks the prefilling stage and KV cache swap-out time, during which the CPU stays idle; 
2) the linear stage of a decoding sub-batch on the GPU is typically much shorter than the attention stage on the CPU, due to the attention's auto-regressive nature and high memory bandwidth consumption. 
As a result, the duration of the attention stage will likely exceed that of the linear stage, causing the CPU to become the bottleneck. 

\item Finally, it is challenging to split two batches evenly in practice. In real-world workloads, it is nearly impossible to ensure that a single batch can be divided into two identical sub-batches, due to different input lengths and unpredictable output lengths. The discrepancies between the batches would likely result in a significant number of idle periods or ``bubbles'' in the pipeline. 
\end{itemize}

\para{Asymmetric pipelining}, as shown in Figure \ref{fig:asymm-itlv}, offers a solution to the aforementioned problems. 
To fully utilize GPU memory, \sysname does \textit{partial offloading}. The KV cache system of \sysname is divided into two separate components: the ``GPU-cache'' located in the GPU's HBM, and the ``CPU-cache'' located in the CPU main memory. For any request that has already been prefilled in the system, its KV cache will either reside entirely in the GPU-cache---designated as a ``GPU-request''---or entirely in the CPU-cache---designated as a ``CPU-request''. Requests are prioritized for storage in the GPU-cache to maximize GPU memory utilization. 

To achieve full GPU-CPU overlapping, \sysname integrates the prefilling stage computation into the GPU decoding sub-batch, so that the prefilling stage computation (in GPU) also happens in parallel with the CPU attention computation. 
This shares a similar spirit as the \textit{selective batching} technique used in Orca~\cite{orca}. 
In our context, this selective batching largely extends the duration of the GPU compute, allowing for a longer overall CPU computing time and enabling more CPU-requests to be incorporated into the batch. 
Further, \sysname leverages the \textit{layer-wise swapping} technique to facilitate the overlapping of KV value transmission with computation. Given that the KV values of newly prefilled requests are computed layer by layer, we can initiate PCIe transmission immediately after each layer's KV value is computed, rather than deferring this process until the end of the entire iteration. 

To simplify the batch division scheme and minimize idle periods, \sysname introduces \textit{asymmetric batch division}. Instead of attempting to create two identical batches,
we consolidate all prefilling requests and GPU decoding requests with a few CPU decoding requests to batch-0, while dispatching the majority of 
CPU decoding requests to batch-1. The two batches are complementary: batch-0 features a long linear stage with little CPU attention computation, whereas batch-1 includes a lengthy attention stage with a very short linear stage. This arrangement not only simplifies system implementation, but also facilitates effective overlapping between the two batches, resulting in an inference iteration characterized by alternative ``long-stages'' and ``short-stages'', rather than uniform stages. 

Asymmetric pipelining offers even more benefits. As shown in Figure~\ref{fig:asymm-itlv}, the non-overlapping segments at the beginning and end are minimized due to this intentional asymmetry. Furthermore, this approach reduces GPU kernel launching overhead, which can be substantial in Python due to its limited multi-threading parallelism, consuming considerable CPU time. In asymmetric pipelining, the GPU attention kernel is only invoked once per iteration, as only one sub-batch will need the GPU attention, compared to twice in symmetric pipelining.

\presub
\subsection{Load-Aware Scheduling}
\postsub
\label{ssec:scheduler}

Real-world inference workloads are complex, irregular, and dynamically changing, \eg, different input/output lengths across requests. 
\sysname scheduler considers asymmetric pipelining and adaptively determines whether the incoming request should be placed on the GPU or CPU, to keep both busy while not overloading anyone. 

\sysname faces a more challenging problem than prior GPU-only inference engines like vLLM. 
First, \sysname needs to form two sub-batches in each iteration, which means we need to consider batch selection and batch splitting at the same time. 
Secondly, blindly putting as many requests as possible into the CPU decoding runqueue does not work well. This is because too many CPU requests might overload the CPU capacity either in memory bandwidth or compute. 
Finally, the two-batch asymmetric pipelining does not always work better than GPU-only inference. 
This is because in asymmetric pipelining, the GPU memory needs to hold two concurrently running sub-batches, and each sub-batch can only achieve half of the maximum batch size; if \sysname cannot offload enough KV caches, the sub-batch size might be even smaller than the GPU-only inference, yielding lower throughput.  

To address these challenges, \sysname follows several principles: 
\begin{itemize}
    \item \textbf{Greedy:} At the beginning of each iteration, \sysname's scheduler would make a GPU-only inference schedule and a two-batch asymmetric pipelining schedule, and would choose the one with higher estimated throughput as the final decision. 
    \item \textbf{Balancing:} For asymmetric pipelining, the scheduler should minimize pipeline ``bubbles''. That is, the estimated CPU busy time and GPU busy time should be as close as possible. 
    \item \textbf{Hiding CPU:} For asymmetric pipelining, there shouldn't be cases when the CPU is busy but the GPU is idle. 
    \item \textbf{Maximizing GPU:} For asymmetric pipelining, in each iteration, the scheduler should pick as many requests as possible from the prefilling waitqueue, and GPU and CPU decoding runqueues (that hold requests running on GPU and CPU respectively). 
\end{itemize}

Concretely, denote each iteration time as $T$ and batch size as $x$, \sysname's goal is to maximize $T/x$. 
As shown in Figure~\ref{fig:asymm-itlv}, $T$ consists of the following components: pre-layer time $T_{prl}$, transformer layer time $T_{tr}$, post-layer time $T_{pol}$, while $T_{tr}$ is the major part that usually takes more than 95\% of time in each iteration. Therefore, we can estimate $T$ as below: 
\begin{align*}
T \approx T_{tr} = L \times (&\max\{T_{po_0} + T_{pr_0}, T_{ca_1}\} + \\
                             & \max\{T_{po_1} + T_{pr_1} + T_{ga_0}, T_{ca_0}\})
\end{align*}
where $T_{po_x}$, $T_{pr_x}$, $T_{ga_x}$, and $T_{ca_x}$ denote time for post-projection, pre-projection, GPU attention, and CPU attention time for batch-$x$, respectively. 
For simplicity, we define $T_{l_x} = T_{po_x} + T_{pr_x}$, which stands for time mainly consumed by multiplying activations with model weights in one transformer layer. 
To estimate $T_l$, $T_{ga}$, and $T_{ca}$, \sysname does offline profiling for typical input/output lengths and uses linear interpolation to approximate the values for other lengths. 
Overall, to stick to principles of ``Balancing'' and ``Hiding GPU'', we would like to guarantee $T_{l_0} \ge T_{ca_1}$ and $T_{l_1} + T_{ga_0}\ge T_{ca_0}$, and minimize the gap between the left-hand side and right-hand side of both inequations. 

Based on the analysis above, we present our load-aware scheduler, with the following procedures on each iteration: 
\begin{enumerate}
    \item \textit{Initialization.} Initialize two empty batch schedules: batch-0 would mostly run on the GPU and contain requests in both prefilling and decoding phases; batch-1 would mostly run on the CPU and only contain requests that calculate decoding attention. 
    \item \textit{Schedule GPU decoding requests.} Try to put every request from the GPU decoding runqueue into batch-0. Then swap out requests until GPU memory can hold all new KV cache, or swap in requests if there is ample space on GPU (Maximizing GPU). 
    \item \textit{Schedule prefilling requests.} Pop the prefilling waitqueue and put requests into batch-0: keep the generated KV cache on GPU if there is enough GPU memory, otherwise, swap out the generated KV cache. Do this repeatedly until the GPU cannot hold the activations in the batch (Maximizing GPU).
    \item \textit{Schedule CPU decoding requests.} Scan the CPU decoding runqueue and put requests into either batch-0 or batch-1, maintaining $T_{ca_0} \le T_{l_1} + T_{ga_0}$ and $T_{ca_1} \le T_{l_0}$. If putting a request into either batch would violate the inequations, skip this request and leave it for the next iteration (Balancing and hiding CPU).
    \item \textit{Reduce prefilling requests.} Remove any prefilling request from batch-0 that would require swapping out the generated KV cache, as long as the above inequations hold. This is to avoid the CPU being idle (Balancing). 
    \item \textit{Make decisions.} Now the two-batch schedule is made; we make the GPU-only schedule by taking batch-0 and excluding all the CPU decoding requests added in step 4. 
    Finally, we compare their estimated $T_{tr}/x$ values and pick the schedule with a larger one (Greed).
\end{enumerate}

\presec
\section{Implementation}
\postsec
\label{sec:implementation}

We implement \sysname based on SwiftLLM~\cite{swiftllm}---a tiny yet powerful LLM inference system for research purposes. 
SwiftLLM achieves vLLM-equivalent (single GPU, default scheduling policy) performance with around 2K lines of code. 
Our \sysname implementation consists of 4K lines of Python and 1.5K lines of C++. 

\para{Efficient CPU Kernels.} We implement a custom C++ torch extension library called Paged-Attention-for-CPU (PACPU) and plug it into SwiftLLM. The PyTorch runtime will directly call a multi-threaded procedure written in C++. Under the hood, PACPU calls yet another extension library written in ISPC~\cite{ispc}---a language for writing SPMD programs on CPUs. ISPC code can be compiled to targets with different sets of vectorized instructions, such as AVX2, AVX512, and ARM Neon, making it perfectly portable between different types of CPUs. 
PACPU utilizes a paged KV cache similar to vLLM with the Paged Attention algorithm to mitigate memory fragmentation. 

\sysname optimizes memory access performance for decoding attention on CPU, as the decoding attention is heavily bounded by memory bandwidth (even with GQA)~\cite{sarathi_server, nanoflow}. 
1) Within a single CPU core, we utilize SIMD memory load/store instructions (features provided by ISPC) to minimize instruction overhead and enhance memory throughput. Additionally, we carefully organize the memory access order to ensure its contiguity.
2) Across different CPU cores, we use a parallelism strategy similar to Flash Decoding~\cite{flash_decoding}. For each request, we partition its computation into individual tasks along the request dimension, allowing each task to access unique and continuous memory at block granularity. These tasks are evenly dispatched to all threads with each thread having an equal number of blocks to process. Finally, we aggregate the partial outputs of each request to obtain the final result. 

\para{Reducing Kernel Launching Overhead.} Due to Python's poor multi-threaded execution performance incurred by its global interpreter lock (GIL), the data plane CPU kernels could not run in parallel with the control plane CUDA kernel launching calls, making kernel launching blocking on the critical path.
To reduce kernel launching overhead, we replaced most of SwiftLLM's Triton-JIT~\cite{triton, triton_oai} kernels with kernels written in CUDA C++. This can effectively reduce the additional kernel selection and launching overhead brought by Triton-JIT. 

\para{Multi-GPU inference.} We redesigned SwiftLLM's architecture to support model sharding and tensor parallelism, as it originally only supported single-GPU inference. 
We utilize Ray actors~\cite{ray} to hold shards of the model and use PyTorch's pre-built communication library (based on NCCL) to handle cross-GPU communication. The underlying mechanism of splitting tensors and collecting results across GPUs is the same as vLLM~\cite{vllm}. 
As a standard approach, each Ray actor also has its partition of CPU KV cache, each responsible for a portion of KV heads to avoid cross-CPU communication. 

\presec
\section{Evaluation}
\postsec
\label{sec:eval}

\subsection{Experiment Setup}
\label{ssec:exp_setup}
\postsub

\textbf{Testbeds.} 
We run our experiments on multiple types of single-GPU instances on the AWS EC2 public cloud, including g5.2/4/8/16xlarge with an A10G GPU and g4dn.4xlarge with a T4 GPU. By default, we use g5.4xlarge for all A10G experiments. 
We also run on an 8$\times$H100 SXM local server to test multi-GPU performance. The specifications of hardware are listed in Table~\ref{tab:hard-spec}. 
Note that the HGX machine has 4 CPU NUMA nodes. We confine our system to running on a single NUMA node (thus 1/4th of the total memory size and bandwidth) when running 2-GPU experiments. 

\begin{table}[ht]
    \centering
    \resizebox{0.95\linewidth}{!}{
    \begin{tabular}{lcccr}
    \toprule
         Name & GPU & CPU (\#cores) & Memory \\
    \midrule
         g5.$n$xlarge & A10G & EPYC 7R32 (2$n$) & 16$n$ GB \\
         g4.4xlarge & T4 & Xeon P-8259CL (8) & 64 GB \\
         HGX & 8$\times$H100 & Xeon 8462Y+ (64) & 2 TB \\
    \bottomrule
    \end{tabular}
    }
    \vspace{-0.05in}
    \caption{Hardware specifications of our testbeds.}
    \label{tab:hard-spec}
\end{table}

\textbf{Models.} 
We evaluate \sysname on the representative LLaMa models including LLaMa-3.1-8B, LLaMa-3.1-70B~\cite{llama3} and LLaMa-2-7B~\cite{llama2}.

\textbf{Baselines.} 
We consider three baselines in our evaluation.

\begin{itemize}
\item vLLM~\cite{vllm} is a popular state-of-the-art LLM inference system. vLLM's default scheduling policy doesn't include selective batching; so we set the \texttt{--enable-chunked-prefill} flag to enable it. 
\item FastDecode~\cite{fastdecode} is an LLM inference system that offloads full decoding attention. Since the original work didn't consider the prefilling stage or implement an end-to-end system, we implemented our own version of FastDecode$^+$. It leverages \sysname's asymmetric pipelining and load-aware scheduling, but offloads all requests' decoding attention to the host CPU. 
\item SwiftLLM~\cite{swiftllm} is a simplified version of vLLM with similar Pythonic implementation and without offloading, upon which \sysname is built. We also make simple modifications to SwiftLLM to support multi-GPU inference. It achieves comparable performance with vLLM on a single GPU, and slightly lower performance than vLLM in 2-GPU settings (\S\ref{ssec:sensitivity}). 
\end{itemize}

\textbf{Workloads.} We use real-world workloads as well as synthetic ones to evaluate our system. 
\begin{itemize}
\item Azure LLM inference trace for coding (AC) \cite{azure_code, splitwise} is a LLM coding trace collected from Azure cloud's production environment. 
\item OpenAI summarization comparison (OSC) \cite{osc} is an open dataset of input text, chosen summary, and rejected summary produced in real-world human chatbot interactions. 
\item Synthetic workloads with various input and output lengths. For a pair of input length $l_i$ and output length $l_o$, we synthesize requests with input and output lengths sampled independently and uniformly from $[0.9l_i, 1.1l_i]$ and $[0.9l_o, 1.1l_o]$, respectively. 
\end{itemize}
We use the AC trace with relatively longer requests on the H100 and A10G GPUs, while using the OSC trace with shorter requests on the lower-end T4 GPU. 

\presub
\subsection{Online Latency vs. Load}
\postsub

\begin{figure*}[!t]
\centering
\begin{minipage}{\textwidth}
    \centering
    \includegraphics[width=0.25\columnwidth]{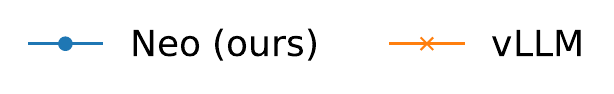}
    \vspace{-1.5mm}
\end{minipage}
\begin{minipage}{0.35\textwidth}
    \centering
    \includegraphics[width=\linewidth]{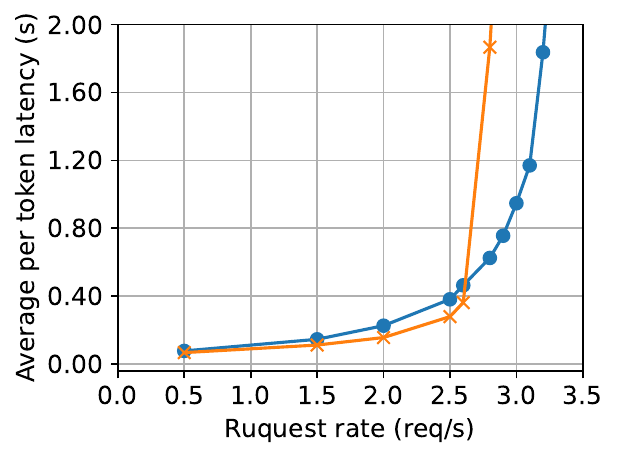}
    \subcaption{2xH100 + LLaMa-3.1-70B + AC.}
    \label{lla}
\end{minipage}
\hfill
\begin{minipage}{0.32\textwidth}
    \centering
    \includegraphics[width=\linewidth]{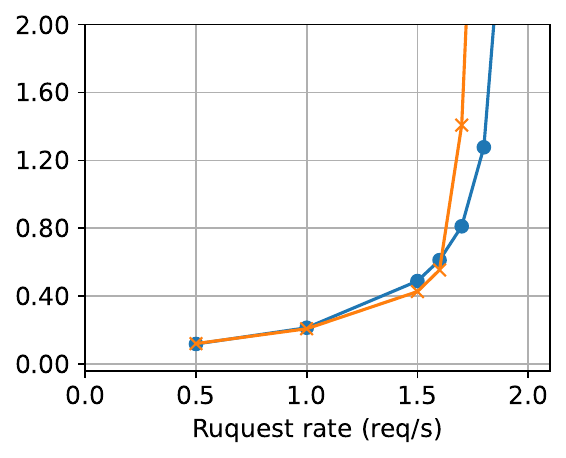}
    \subcaption{A10G + LLaMa-3.1-8B + AC.}
    \label{llb}
\end{minipage}
\hfill
\begin{minipage}{0.32\textwidth}
    \centering
    \includegraphics[width=\linewidth]{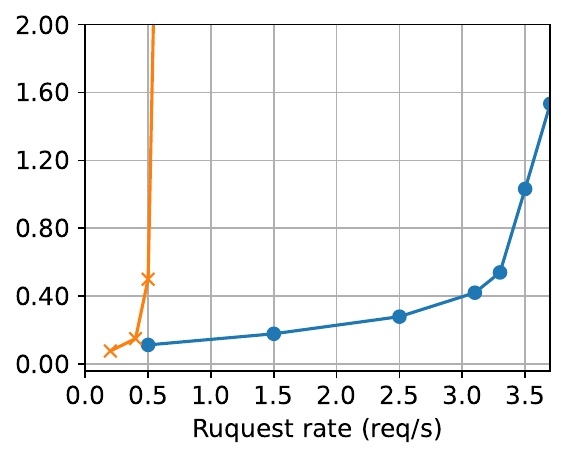}
    \subcaption{T4 + LLaMa-2-7B + OSC.}
    \label{llc}
\end{minipage}
\postfig
\caption{Load-latency curve comparison between \sysname and vLLM on 3 different settings. For each request, we compute its per-token latency by dividing its full latency by its output token number, and then we take the average among all requests.}
\postfigcaption
\label{fig:load-latency}
\end{figure*}

\begin{figure*}[!t]
\begin{minipage}{0.33\textwidth}
    \centering
    \includegraphics[width=\linewidth]{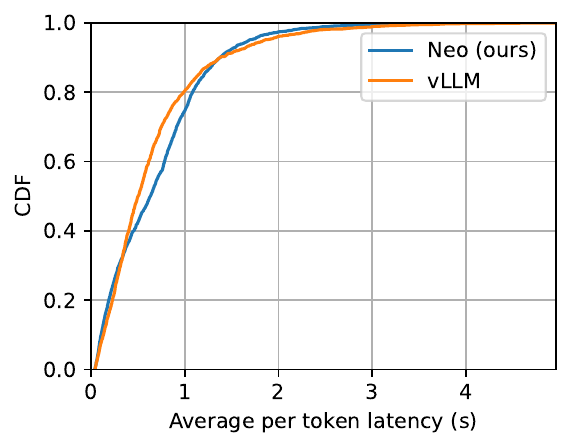}
    \postfig
    \caption{Latency distributions of \sysname and vLLM in A10G+LLaMa-3.1-8B+AC setting at request rate of 1.6/s. The distributions are both skewed because of the skewed request length distribution of the dataset.}
    \postfigcaption
    \label{fig:lat-dist}
\end{minipage}
\hfill
\begin{minipage}{0.65\textwidth}
    \centering
    \begin{minipage}{\textwidth}
        \centering
        \includegraphics[width=0.6\columnwidth]{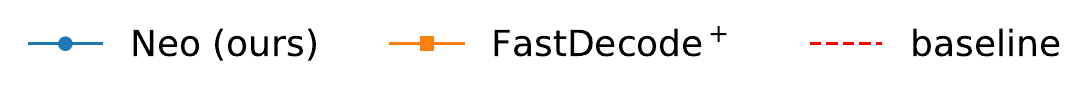}
        \vspace{-1.5mm}
    \end{minipage}
    \begin{minipage}{0.495\textwidth}
        \centering
        \includegraphics[width=\linewidth]{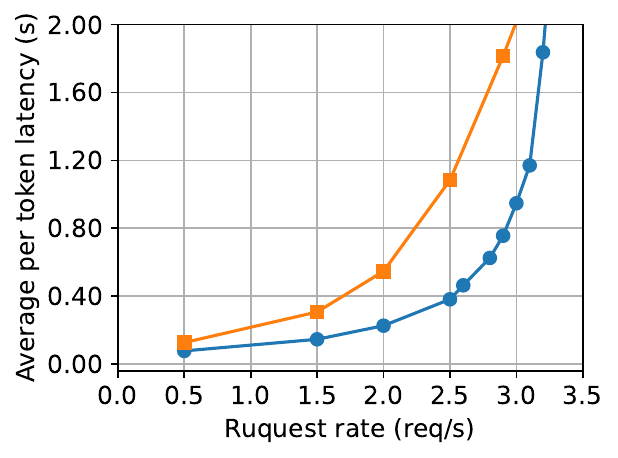}
        \subcaption{Online latency.}
        \label{cfa}
    \end{minipage}
    \hfill
    \begin{minipage}{0.495\textwidth}
        \centering
        \includegraphics[width=\linewidth]{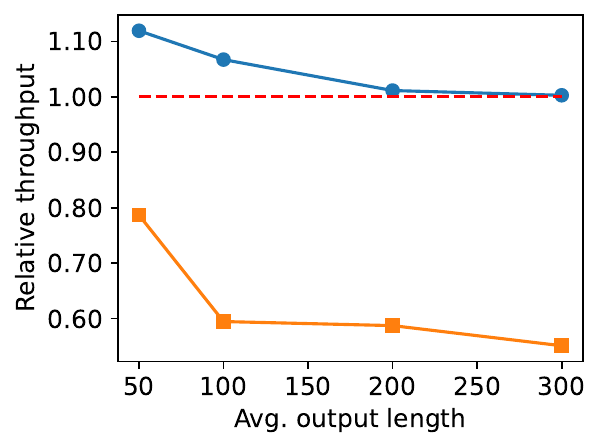}
        \subcaption{Offline throughput.}
        \label{cfb}
    \end{minipage}
    \postfig
    \caption{Comparison of \sysname and FastDecode$^+$ in 2xH100+LLaMa-3.1-70B setting. Figure (a) compares their latency on the AC dataset. Figure (b) shows their relative throughput to the baseline, where we fix the input length to 2000 and vary the output length. We use the GPU-only version of \sysname (\ie, SwiftLLM) as the baseline.}
    \postfigcaption
    \label{fig:comp-fsdc}
\end{minipage}
\end{figure*}

We evaluate the online latency
of \sysname under various request rates. We sample request arrival timestamps following the Poisson process. 
As Figure \ref{fig:load-latency} shows, \sysname sustains higher loads than vLLM in all listed hardware/model settings while providing comparable latencies at low rates. 
\sysname achieves 14.3\% higher throughput on H100 (at 2 sec latency), 6.40\% higher on A10G (at 2 sec latency), and 563\% higher on T4 (at 1 sec latency). 
In the T4+LLaMa-2-7B setting, we achieve nearly 6$\times$ throughput gains over vLLM; this is because the T4 GPU
has an extremely constrained memory budget for the KV cache, severely limiting vLLM batch size, while \sysname could offload the KV cache to the CPU, achieving a much larger batch size. 

At extremely low request rates, \sysname's latency behaves exactly the same as vLLM. As the request rate grows, \sysname's latency gets slightly higher for the following reasons: 1) vLLM is heavily optimized, while \sysname features less engineering optimizations for simplicity and performance clarity---this is especially true in multi-GPU settings. 2) \sysname actively seeks opportunities to offload requests, even though offloading may not help, which consequently leads to more system-level overheads in scheduling and swapping.

Figure~\ref{fig:lat-dist} further compares the latency distributions of \sysname and vLLM, and shows that our throughput gains do not come at the cost of latency. Our inference latency is comparable to vLLM's at all percentages. 

\begin{figure*}[!t]
\centering
\begin{minipage}{0.34\textwidth}
    \centering
    \includegraphics[width=\linewidth]{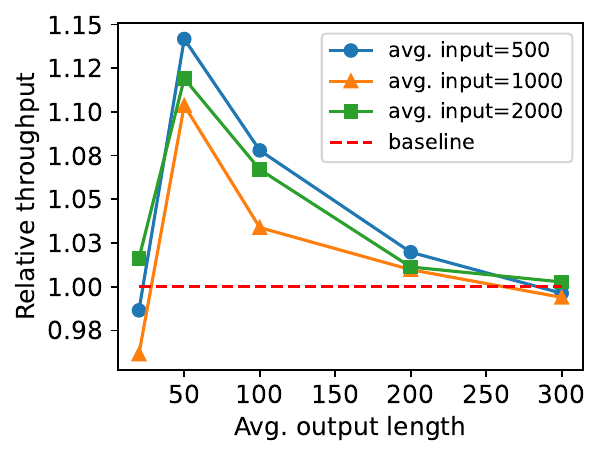}
    \subcaption{2xH100 + LLaMa-3.1-70B.}
    \label{psa}
\end{minipage}
\hfill
\begin{minipage}{0.32\textwidth}
    \centering
    \includegraphics[width=\linewidth]{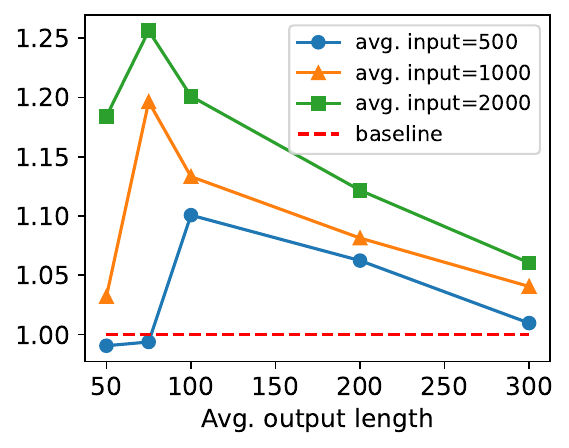}
    \subcaption{A10G + LLaMa-3.1-8B.}
    \label{psb}
\end{minipage}
\hfill
\begin{minipage}{0.32\textwidth}
    \centering
    \includegraphics[width=\linewidth]{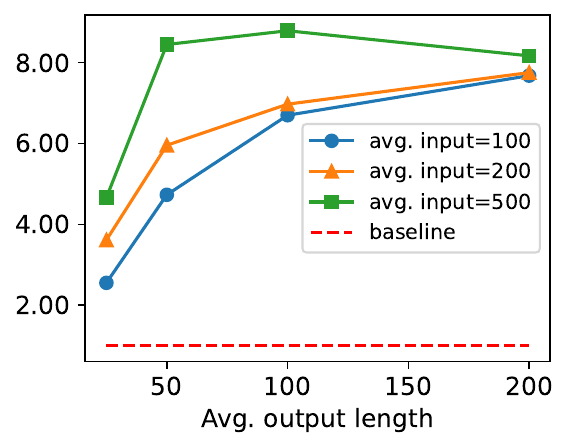}
    \subcaption{T4 + LLaMa-2-7B.}
    \label{psc}
\end{minipage}
\postfig
\caption{Relative throughput in different settings and different synthetic workloads, with GPU-only \sysname (\ie, SwiftLLM) as the baseline.}
\postfigcaption
\label{fig:param-sweep}
\end{figure*}

\begin{figure*}[!t]
\centering
\begin{minipage}{0.33\textwidth}
    \centering
    \includegraphics[width=\linewidth]{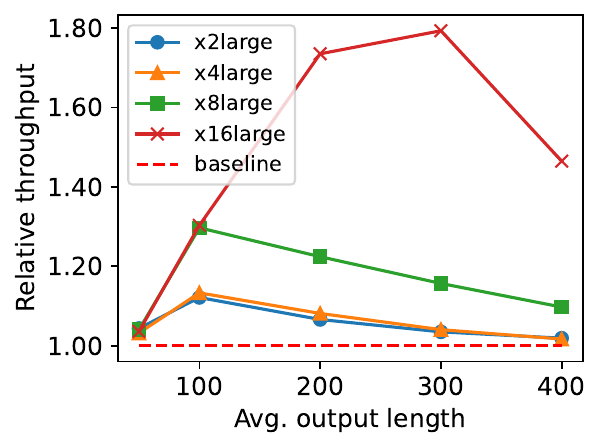}
    \subcaption{CPU sensitivity.}
    \label{fig:cpu-sen}
\end{minipage}
\hspace{0.08\textwidth}
\begin{minipage}{0.33\textwidth}
    \centering
    \includegraphics[width=\linewidth]{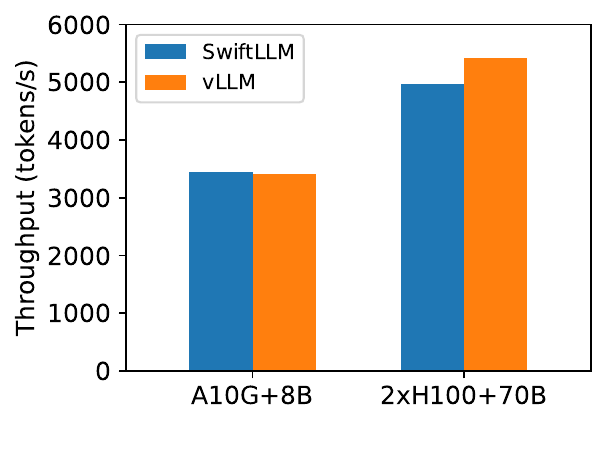}
    \subcaption{SwiftLLM vs. vLLM.} 
    \label{fig:vvb}
\end{minipage}
\postfig
\caption{Sensitivity study. Figure (a) shows the relative throughput of \sysname on different AWS EC2 \texttt{g5} instances, with GPU-only \sysname (\ie, SwiftLLM) as the baseline.
Figure (b) compares SwiftLLM's throughput on the AC dataset with vLLM in both multiple and single GPU settings. Note that we implement the multi-GPU support for SwiftLLM (\S\ref{sec:implementation}). }
\postfigcaption
\label{fig:sensitivity}
\end{figure*}

\presub
\subsection{Comparison with FastDecode$^+$}
\postsub

\begin{figure*}[!t]
\centering
\end{figure*}

We compare \sysname and FastDecode$^+$ in terms of both latency and throughput. The results are shown in Figure~\ref{fig:comp-fsdc}.
FastDecode$^+$'s higher latency is caused by its inflexibility to tackle irregular workloads. 
When there are many requests in the CPU decoding runqueue but few or no requests in the prefilling waitqueue, FastDecode$^+$ would have no choice but to launch CPU batches to make progress, hindering overall performance, while \sysname could simply launch GPU batches to utilize GPU resources. 

Furthermore, \sysname is always better in throughput than baseline because its scheduler can always decide to fall back to GPU-only mode. However, as output length grows, FastDecode$^+$ becomes CPU bounded, and its performance drops quickly to less than 60\% of baseline's performance.

\presub
\subsection{Varying Input/Output Lengths}
\postsub

We further examine \sysname's performance over various workloads with different input/output lengths. 
As shown in Figure~\ref{fig:param-sweep}, where we fix input length and tweak output length, \sysname achieve up to 14\%, 26\%, and 750\% throughput gains on H100, A10G, and T4, respectively. 
When the output length is short, \sysname may perform slightly worse than the baseline because \sysname still attempts to put some requests onto the CPU and swap them back in later, incurring slight overheads. 
As the output length increases, \sysname's gains first grow to the maximum point, where GPU and CPU times are exactly balanced, then gradually drop as the system launches a larger proportion of GPU-only batches. \sysname performance would be close to the baseline when output length gets large enough, sometimes slightly worse due to suboptimal scheduling decisions caused by the inevitable inaccuracy of the offline performance profiling. 

\presub
\subsection{Sensitivity Study}
\postsub
\label{ssec:sensitivity}

\para{Impact of CPU capacity.} We first study the impact of CPU capacity on throughput gains. To examine the impact of CPU memory bandwidth, we use 4 kinds of AWS EC2 instances, namely g5.2xlarge, g5.4xlarge (what we used in previous experiments), g5.8xlarge, and g5.16xlarge. These instances all have 1 A10G GPU and thus the same baseline (non-offloading) performance. 
However, their offloading performance varies due to different CPU cores, memory sizes, and memory bandwidths. 
A g5.$n$xlarge generally has $2n$ CPU cores (\ie, $4n$ hyperthreads)
and $16n$ GB CPU memory; g5.2xlarge has nearly the same peak memory bandwidth as g5.4xlarge, while g5.8xlarge has about twice the bandwidth of g5.4xlarge, and g5.16xlarge has about twice the bandwidth of g5.8xlarge. In experiments, we set the CPU KV cache size proportionally to the number of cores. 
Figure \ref{fig:cpu-sen} shows the results. 
\sysname achieves up to 12.2\%, 13.3\%, 29.7\%, and 79.3\% higher throughput over the baseline under different CPU capacities. 
When the output length is short, these instances have nearly the same throughput gains as the workload mainly runs on the GPU. 
As output length increases, the instances with less CPU memory bandwidth start to drop performance earlier. The peak throughput gain is positively related to the CPU memory bandwidth. 
This supports the fact that the memory bandwidth, rather than computing power (\ie, number of cores), is the factor that determines the performance of attention operation on CPUs. 

\para{SwiftLLM vs. vLLM.}  We now examine the gap between our baseline system and vLLM. We feed the Azure Code trace all at once to both systems and examine the GPU token throughput, \ie, the total time elapsed divided by the total number of tokens processed (input length + output length). We evaluate in both single-GPU (A10G + LLaMa-3.1-8B) and multi-GPU (2xH100 + LLaMa-3.1-80B) settings.

Figure \ref{fig:vvb} shows the results. As SwiftLLM is initially targeted at single-GPU inference and mimics vLLM implementation, it achieves comparable throughput with vLLM in a single A10G + LLaMa-3-8B setting. However, SwiftLLM is slightly worse than vLLM (8.8\% lower throughput) in the 2-GPU setting; this is due to our less optimized tensor parallelism implementation compared to production-grade vLLM. 
We leave as future work integrating \sysname into vLLM to measure our 4-GPU and 8-GPU performance gains. 

\presec
\section{Discussion}
\postsec

\para{Compare to chunked-prefill.} 
\sysname's performance benefits are essentially two-fold: 1) larger GPU batch size, and 2) shifting some unbatchable decoding attention operations to CPU, thus making up room on GPU for other batchable operations. 
The second benefit source shares a similar spirit as the chunked-prefill technique in Sarathi-Serve~\cite{sarathi_server}, which breaks a prompt into multiple chunks to launch more prefill-decode mixed batches and thus less decoding-only batches (that contain unbatchable decoding attention operations). 

However, chunked-prefill suffers from several drawbacks that \sysname does not have. 
First, chunked-prefill consumes significantly more GPU memory bandwidth, because the KV cache of all previous chunks needs to be loaded repetitively to compute for the subsequent chunk~\cite{distserve}. 
Second, chunking-prefill would not work well on memory-constrained GPUs, because the resulting small batch size would limit the opportunity of piggybacking decode on prefill chunks while saturating the GPU. 
In contrast, \sysname relies on large CPU memory to ensure throughput gains. 
Nevertheless, we believe \sysname could integrate with chunked-prefill, providing a larger design space. We leave it for future work. 

\para{Offloading other parts to CPU.} \sysname implicitly assumes putting all model weights on the GPU and only offloading attention computation to the CPU is the most efficient way to balance GPU and GPU loads. However, in prior work like \cite{flexgen} and \cite{powerinfer}, people offload components other than attention to the CPU. This is not only useful when the GPU's memory is too constrained to hold all model weights, but also potentially beneficial even if the GPU could hold all the model weights. For example, when requests in the workload have too few output tokens, \sysname would be bounded by GPU computation, while the CPU is mostly idle. Therefore, offloading some of the dense operations to the CPU could alleviate the GPU's pressure in these extreme workloads. Nevertheless, the actual gain needs to be validated by further exploration.

\para{\sysname usage scenarios.} \sysname works best in scenarios where the GPU memory is constrained such that it limits the batch size and underutilizes the GPU compute. 
These scenarios will likely hold for a long time, as the GPU compute capacity continues growing while its memory size stays relatively stagnant, \eg, H100 triples the compute of A100 but with the same 80GB memory size~\cite{a100_datasheet, h100_datasheet}. 
\sysname will degrade to non-offloading mode when the GPU has enough memory to reach a batch size that can saturate the GPU. 
Another usage scenario of \sysname is the economic serving of LLM models by leveraging cheap and abundant CPU resources that already exist in current datacenters. 

\para{Using remote CPUs.} \sysname focuses on improving inference throughput by only using the host CPU. In order to gain more throughput, \sysname could be extended to support remote CPU workers. However, CPU memory bandwidth in current commercial clouds is still expensive, and as \cite{fastdecode} shows, cross-machine transfer latency could be yet another bottleneck. We leave this for future work. 

\presec
\section{Related Work}
\postsec

\para{GPU-efficient LLM inference. } 
There is a line of work on optimizing the efficiency of LLM inference on GPUs, 
including general inference systems from Orca~\cite{orca}, vLLM~\cite{vllm}, SGLang~\cite{sglang}, FastServe~\cite{fastserve}, Sarathi-Serve~\cite{sarathi_server}, NanoFlow~\cite{nanoflow}, DistServe~\cite{distserve}, and more~\cite{splitwise, dejavu, no_interference}, 
and low-level GPU kernel optimizations from 
FlashDecoding~\cite{flash_decoding}, FlashDecoding++~\cite{flashdecoding++}, and FlashInfer~\cite{flashinfer}.
\sysname leverages several techniques from these work such as selective batching~\cite{orca} and paged attention~\cite{vllm}, and could be used in parallel with others, \eg, leveraging \sysname to optimize the decoding phase in DistServe~\cite{distserve}. 
Another line of work leverages sparcification and quantization techniques to trade accuracy for performance, including AWQ~\cite{awq}, SparseGPT~\cite{sparsegpt}, AlphaTuning~\cite{alphatuning}, GPT3.int8()~\cite{gpt3int8}, GPTQ~\cite{gptq}, ZeroQuant~\cite{zeroquant}, SmoothQuant~\cite{smoothquant}, StreamingLLM~\cite{attention_sink}, and more~\cite{longformer, qbert, sparsity_in_dl}. 
Different from them, \sysname does not compromise accuracy. 

\para{Offloading for LLM serving. }
Many existing work offloads LLM models, activations, KV cache, or computations to the CPU for offline scenarios that trade latency for throughput, such as FlexGen~\cite{flexgen}, HeteGen~\cite{hetegen}, PowerInfer~\cite{powerinfer}, and TwinPilots~\cite{twinpilots}. 
InstInfer~\cite{instinfer} further offloads to computational SSD to lower the inference cost. 
FastDecode~\cite{fastdecode} targets similar online scenarios as \sysname, but lacks critical designs to address the load imbalance between the GPU and CPU (see \S\ref{introduction}). 

\presec
\section{Conclusion}
\postsec

\sysname is a CPU offloading system for online LLM inference to increase GPU batch sizes and improve inference throughput. 
It features asymmetric pipelining and load-aware scheduling to fully leverage both GPU and CPU resources without overloading them. 
\sysname achieves up to 14\%-7.5$\times$ (depending on GPUs) higher throughput than GPU-only inference systems across a variety of workloads and model sizes, while maintaining the same latency. 
We will open source \sysname codebase to encourage more research on cost-efficient LLM inference.

\bibliography{ref}
\bibliographystyle{mlsys2025}

\end{document}